**Does the Müller-Lyer illusion induced by a goalkeeper configuration influence soccer penalty kicks?**

Sufiaan Ahmed, Tyrese Lindsay,

& James W. Roberts*

Liverpool John Moores University, Brain & Behaviour Research Group,

Research Institute of Sport & Exercise Sciences (RISES),

Tom Reilly Building, Byrom Street, Liverpool, UK, L3 5AF.

**\*Corresponding author**

James W. Roberts

Liverpool John Moores University, Brain & Behaviour Research Group,

Research Institute of Sport & Exercise Sciences (RISES),

Byrom Street, Tom Reilly Building, Liverpool, UK, L3 5AF

E-mail: J.W.Roberts@ljmu.ac.uk

**Title:** Does the Müller-Lyer illusion induced by a goalkeeper configuration influence soccer penalty kicks?

**Abstract:** In soccer penalty kicks, goalkeepers that orient their arms upward compared to downward can be misperceived as being taller─effectively recreating the Müller-Lyer illusion. The present study elaborates on previous research surrounding a potential illusion-induced bias in penalty kicks. Participants were exposed to goalkeeper configurations within a virtual goal including arms-parallel, arms-down, arms-out and arms-up. They separately judged the perceived size of the goalkeeper, and executed penalty kicks. The perceived size was near fully consistent with the intended illusion. Meanwhile, the penalty kicks indicated wider a horizontal position following arms-out, and lower vertical position following arms-up. Likewise, there was no relation between the biases expressed in perception and action. While goalkeepers can elicit a perceptual illusion, this does not extend to influencing the penalty kick itself. Instead, other contextual cues appeared more relevant including the proximity between the goalkeeper and goalposts, and with it, the available space in the goal.

**Titre:** L'illusion de Müller-Lyer induite par une configuration de gardien de but influence-t-elle les tirs au but au football?

**Résumé:** Lors des penaltys de football, les gardiens qui orientent leurs bras vers le haut plutôt que vers le bas peuvent être perçus à tort comme plus grands, recréant ainsi l'illusion de Müller-Lyer. La présente étude s'appuie sur des recherches antérieures concernant un biais potentiel induit par l'illusion lors des penaltys. Les participants ont été exposés à différentes configurations de gardiens dans un but virtuel, incluant bras parallèles, bras baissés, bras tendus et bras levés. Ils ont évalué séparément la taille perçue du gardien et ont exécuté les penaltys. La taille perçue correspondait presque parfaitement à l'illusion recherchée. Parallèlement, les penaltys indiquaient une position horizontale plus large après bras tendus, et une position verticale plus basse après bras levés. De même, aucune relation n'a été observée entre les biais exprimés dans la perception et l'action. Si les gardiens peuvent susciter une illusion perceptuelle, celle-ci n'influence pas le penalty lui-même. En revanche, d'autres indices contextuels sont apparus plus pertinents, notamment la proximité entre le gardien et les poteaux, et donc l'espace disponible dans le but.

**Mots clés:** visée à distance; perception; voies visuelles; allocentrique

**Introduction**

The penalty kick within soccer is often recognised as one of the most critical moments within sport as it can provide an exponentially higher chance of scoring whilst outside of open play, and ultimately determines the outcome of competitive matches when it comes to shootouts. Consequently, it is a setting that is most susceptible to performance-related stress (Jordet et al., 2007); and hence goalkeepers may adopt strategies in advance of the kick itself in order to give themselves an advantage and hinder the opponent kicker (e.g., imperceptible off-centre positioning: Masters et al., 2007, arm waving: Wood & Wilson, 2010a; Furley et al., 2017; for a review, see Memmert et al., 2013).

Along these lines, van der Kamp and Masters (2008) made a rather interesting observation on the typical arm location of goalkeepers when they are faced with a penalty kick in a game. That is, depending on how far up or down their arms are located, goalkeepers may effectively recreate an amputated version of the empirically tested perceptual illusion known otherwise as the *Müller-Lyer illusion*. This illusion involves the perceived size of a single vertical shaft being made to look longer or shorter having merely intersected with a set of tails that are oriented at obtuse (135°) (Y) and acute (45°) (λ) angles, respectively. In the context of goalkeepers, they may be perceptibly taller or smaller having raised and lowered their arms, respectively. As a result, it was found that these sorts of configurations could influence the horizontal position of overarm throws at goal in a way that was somewhat consistent with the direction of the perceived size of the goalkeeper. In a similar vein, Shim et al. (2014) found an similar illusion-induced bias, but only when the throws were made to the perceived reach (i.e., "throw directly to the hand or where the hand could reach") (Exp. 1) or height (i.e., "throw to the head as if the goalkeeper were to dive for the ball") (Exp. 2) of the goalkeeper.

However, there have been some discrepancies in the report of illusory effects from the fore mentioned studies, which can make things rather difficult when trying to formulate any definitive conclusions and subsequent recommendations. Namely, the so-called illusion-induced bias was evident only for some of the configurations (e.g., arms-up > arms-out, arms-up ≈ arms-down; van der Kamp & Masters, 2008), while it became completely nullified when aiming to (rather than in reach of) the goalkeeper (Exp. 4; Shim et al., 2014). Additionally, there was also an effect of the configurations on the vertical position of throws that was almost entirely independent of the illusion, where the ball was thrown lower for the arms-up and arms-out compared to arms-down and arms-parallel (van der Kamp & Masters, 2008).

With this in mind, there have also been somewhat mixed findings within the literature around the (lack of) illusory effects on other far-aiming tasks. Namely, in a shuffleboard-type sliding task, there was report of an illusion-induced bias involving the Judd illusion (i.e., perceived mid-point of a horizontal line is off-set toward the right or left when there are leftward and rightward facing arrows at the endpoints, respectively) (van der Kamp et al., 2009). Alternatively, in a manual hitting action intended to propel a ball into the distance toward a Müller-Lyer configuration, there was a limited illusion-induced bias at least when the configuration remained stationary or unchanged (Caljouw et al., 2010, 2011) (see also, Witt et al., 2012; Wood et al., 2013).

Broadly speaking, these discrepancies may be attributed to the context of the related tasks, and their pre-occupation with the functionally distinct neural pathways involving vision-for-perception and vision-for-action (Milner & Goodale, 1995, 2008). Here, it is suggested that the ventral pathway (comprising the occipitotemporal cortex) computes the relative metrics (allocentric) to give rise to an illusion mostly within perception (e.g., tails-in < tails-out Müller-Lyer configuration), while the dorsal pathway (comprising the posterior parietal cortex) computes the absolute metrics (egocentric) to effectively nullify the illusion solely within action (e.g., tails-in ≈ tails-out Müller-Lyer configuration) (see also, Glover, 2004). This logic may explain how illusion-induced biases are more heavily inflicted for pre-planned far-aiming tasks as opposed to visually-guided near-aiming tasks (Bruno et al., 2008; cf. Elliott et al., 2010; Mendoza et al., 2005), as well as movements in the absence compared to presence of visual sensory feedback (Elliott & Lee, 1995; Gentilucci et al., 1996; Westwood & Goodale, 2003; see also, Glover & Dixon, 2001, 2002).

At this juncture, it is of interest to more closely consider the potential for illusion-induced biases caused by goalkeeper configurations. While the previous related studies that examined illusory effects of goalkeeper configurations used a ball-throwing task either with (van der Kamp & Masters, 2008) or without a goal (Shim et al., 2014), the present study alternatively examines illusory effects within penalty kicks. Additionally, in order to assume a genuine illusion-induced bias, it is suggested that there ought to be a close relation between the direction of perception and action (Knol et al., 2017; Smeets et al., 2020; see also, Roberts et al., 2021). Hence, we also examined the potential covariation between the perceived size of the goalkeeper and illusory effects within penalty kicks.

Participants separately judged the perceived size of, and executed penalty kicks toward, a virtual goalkeeper in the centre of a goal that assumed different configurations including arms-parallel (0°), arms-down (45°), arms-out (90°) and arms-up (135°). Based on the previous suggestions, we predict that the horizontal position of the ball would be furthest away from the goalkeeper for the perceptually larger arms-up configuration, although it would be closest for the perceptually smaller arms-down configuration. What's more, we predict that these illusory effects would positively covary between perception and action.

**Method**

*Participants*

An apriori power analysis was conducted using G*Power software (v. 3.1.9.4) including the following input parameters: *α* = .05, 1-*β* = .80, 4 levels of measurement, *f* = .40 / $\eta_p^2$ = .14 (large; Cohen, 1988) (based on previous findings of the perceptual illusions within far aiming tasks; e.g., van der Kamp et al., 2009; Shim et al., 2014). The subsequent estimated number was 10 participants.

The present study recruited 11 male participants (*M* age = 20.6 years, *SD* = 1.3) with a minimum of 3 years of competitive experience in soccer. All participants reported being free of any musculoskeletal injury, and perceptual or neurological condition. The study was approved by the local institutional research ethics committee, and designed and conducted in accordance with the 1964 Declaration of Helsinki.

*Stimuli*

Stimuli were displayed on a large projector screen (4.14 x 3.30 m, 1280 x 1024 pix resolution (5:4 aspect ratio)) via a remote computer using an in-house designed Matlab routine (v. 2022a) (MathWorks, Natick, MA) running Psychtoolbox (v. 3.0.18.13) (Pelli, 1997).

A virtual goal was used for the experimental stimuli comprising of goalposts and a naturalistic background. While these features may subtly influence perception and/or action independent of goalkeeper configuration, they remained constant throughout the study so that any subsequent effects could most likely be attributed to changes in the configurations themselves. Each of the goalkeeper configurations involved creating a life-size image of a goalkeeper that was pre-experimentally photographed within the lab. Here, the goalkeeper stood upright at centre with their feet near shoulder-width apart, and both arms symmetrically fanned out to different locations depending on the intended configuration. Specifically, the arms were located at near 0°, 45°, 90° and 135° with respect to the upper-body to assume the arms-parallel, arms-down, arms-out and arms-up configurations, respectively (see Figure 1). In this regard, the arms-down and arms-up configurations were intended to be perceptually smaller and larger, respectively. Meanwhile, the arms-out configuration was deemed to uphold a potential influence of the arms independent of any illusory context, and the arms-parallel configuration provided an experimental control that was altogether independent of both the illusory context and arms.

Meanwhile, for a separate perceptual illusion task (see later under *Task and Procedures*), a virtual black line (5-pix width) was drawn vertically on a white background. The location (x-axis) and initial size (y-axis) of the line would vary between presentations with the option to adjust the size based on participant responses.

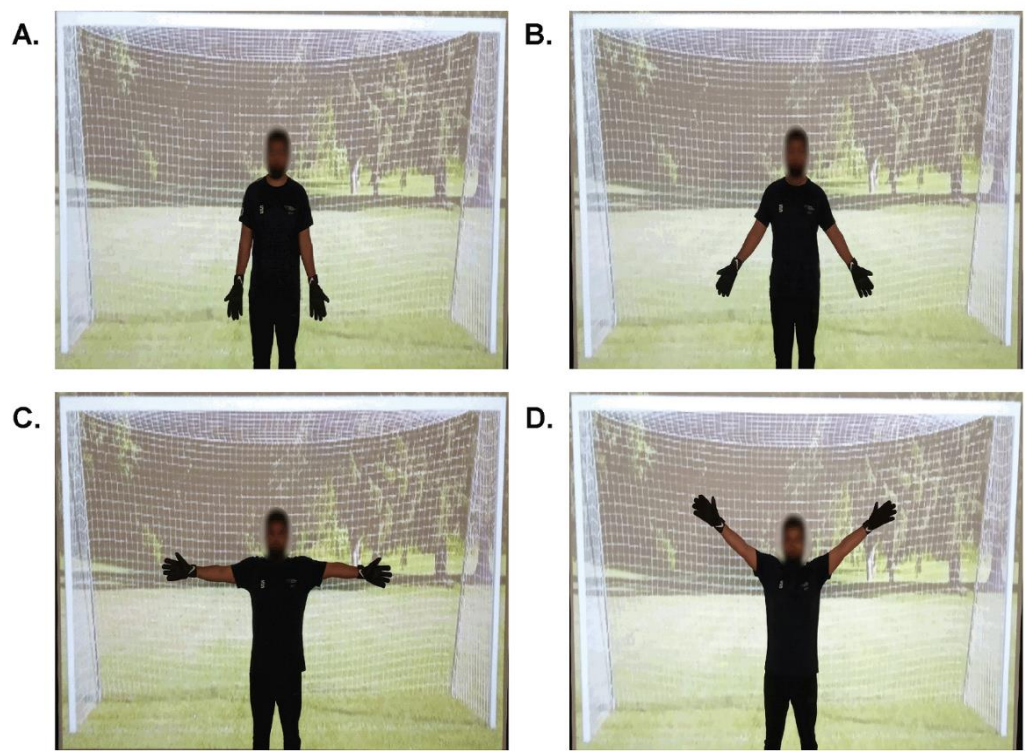

Figure 1. Goalkeeper configurations including arms-parallel (A), arms-down (B), arms-out (C) and arms-up (D).

*Task and Procedure*

The goalkeeper stimuli were initially presented for a period of 3 s, which corresponds with the typical time range for preparing a penalty kick in soccer (Jordet et al., 2009; Wood & Wilson, 2010a). Participants stood or took penalty kicks at 6.22 m from the screen with a view to it being approximately equivalent to the visual angle of the goal within real-life penalty kicks (18.4°).

Participants were assigned both perceptual illusion and penalty kick tasks. For the perceptual illusion task, a virtual line was presented subject to it being adjusted in size until reaching the same size as the goalkeeper height from the previous stimulus. This adjustment was made by using a keypad that was connected to the computer via a Universal Serial Bus (USB) extension. Here, the size of the line was either increased or decreased by ~5 pix (1.6 cm) each time that the up and down arrow keys were pressed, respectively. The final size was confirmed by pressing on the 'enter' / 'return' key in a self-selected time (grand $M$ initiation time = 6.57 s, SD = 2.58).[1]

For the penalty kick task, participants were simply instructed to take a penalty kick as they would within a real game. That is, they would initially step back to their preferred distance (approx. 2-5 m) and look toward goal in preparation for the kick. Therein, they would step or run up to kick the ball at goal while avoiding the goalkeeper. The goalkeeper had to be treated as per a typical penalty kick; that is, with the capacity to intercept or save the ball. A regulation-sized sponge ball was kicked for safe indoor lab use, which was sufficient for the present study purposes due to the short distance to the target and continued sensitivity to influences on perception including potential illusory effects. The kicks were recorded using as a digital camera (Canon Legria FS200, London, UK) (720 x 576 pix spatial resolution, 25 Hz temporal resolution) that was mounted on a tripod at a height of ~3.5-m, and located directly in line and behind the ball at ~10 m from the screen.

---

[1] One-way repeated-measures ANOVA on initiation times within the perceptual task indicated a main effect of configuration that only approached conventional levels of significance, $F(3,30) = 2.86$, $p = .053$, $\eta_p^2 = .22$. Polynomial contrasts indicated no significant linear, $F(1,10) = .04$, $p = .85$, $\eta_p^2 = .004$, nor quadratric, $F(1,10) = 2.32$, $p = .16$, $\eta_p^2 = .19$, trend, while only the cubic trend approached significance, $F(1,10) = 4.19$, $p = .068$, $\eta_p^2 = .30$, where the arms-parallel took the longest time to respond. Meanwhile, there was no significant difference between the arms-down and arms-up, $t(10) = 1.15$, $p = .28$, $d_z = .35$. Thus, the time it took for participants to respond did not seem to confound their perception of, nor action toward, the configurations.

The order of the tasks was counter-balanced between participants. Each task comprised a total of 20 trials including 5 trials for each of the possible goalkeeper configurations. The trials were presented at random with the caveat that a configuration could not reappear until the remaining other configurations appeared the same number of times (i.e., once every 5 trials). In addition, there was an initial practice/familiarisation of the penalty kick task involving a single run of each of the possible configurations. The study took near 60 min for each participant to complete.

*Dependent Measures and Data Analysis*

The pixelated size of the final selected figures from the perceptual illusion task were automatically stored within Matlab, and later converted to centimetres (3.09 pix/cm). Therein, we normalized these values by calculating a ratio between the selected and veridical size of the virtual goalkeeper (<1 indicating a smaller estimate than the veridical size).

Digital recordings of the penalty kicks were analysed frame-by-frame using Kinovea software (v.0.9.5). The moment of interest was identified as the first frame that the ball had finally reached the screen with the virtual goal. The ball position was calculated by drawing calibrated virtual lines between the ball and pre-allocated reference points; that is, the goalkeeper mid-line and head-height for horizontal and vertical positions, respectively. Specifically, the horizontal position was taken as the absolute (unsigned) distance between the ball and mid-line with no indication of direction (i.e., left/right) because there was no hypothesized bias within this particular plane. Meanwhile, the vertical position was taken as the signed distance between the ball and head-height with a close indication of direction (i.e., up/down) because it was alternatively linked to a hypothesized bias.

Any trials that missed the target were still coded providing they continued to feature within the frame of the digital recording. Upon review, there were no trials where the ball had reached outside of the frame, which meant all possible trials was permissible for eventual analysis. Participant mean values were calculated for each of the fore mentioned measures. These data were then entered separately into a one-way repeated-measures ANOVA. The Sphericity assumption (i.e., equal variance of differences) was evaluated using Mauchly's test, and it was deemed to be violated if the test reached significance ($p < .05$). In the event of a violation, then the Greenhouse-Geisser corrected value was adopted if Epsilon ($\varepsilon$) was <.75, although the Huynh-Feldt corrected value was adopted if $\varepsilon$ was alternatively ≥.75 (N.B., original Sphericity-assumed or uncorrected degrees-of-freedom are reported). The effect size measure of interest was partial eta-squared ($\eta_p^2$). Subsequently, we ran follow-up non-orthogonal polynomial contrasts with the order of the different levels of the independent variable being entered according to our hypothesized direction of effects from the smallest to largest extent: 1) arms-down, 2) arms-parallel, 3) arms-out, 4) arms-up (van der Kamp & Masters, 2008; Shim et al., 2014).

In order to corroborate the aforementioned analyses, and based on the most critical comparison of interest from our apriori hypotheses, we ran an additional paired-samples t-test (uncorrected) to compare arms-down and arms-up, and using Cohen's $d_z$ as an effect size measure. Along these lines, in order to more directly evaluate the relation between perceived size and ball position, we also ran bivariate Pearson correlations between the normalized perceived size and ball position differences in arms-up and arms-down separately for the horizontal and vertical dimensions. Parametric statistical assumptions were checked in advance using a combination of the Shapiro-Wilk test, and frequency-distribution and Q-Q plots with each of the repeated-measures levels subsequently indicating normality. Significance was declared at $p < .05$ for all statistical tests.

**Results**

For the normalized perceived size, there was a significant effect of configuration, $F(3,30) = 8.07$, $p = .006$, $\eta_p^2 = .45$. Subsequent polynomial contrasts indicated a significant linear, $F(1,10) = 9.04$, $p = .013$, $\eta_p^2 = .48$, and quadratic, $F(1,10) = 10.54$, $p = .009$, $\eta_p^2 = .51$, component, where the estimated size remained relatively unchanged between arms-down and arms-parallel, but then grew increasingly larger with each subsequent level of configuration (i.e., arms-out to arms-up) (see Figure 2). Consistent with the illusory context, the further t-test indicated a significantly smaller estimated size for the arms-down compared to arms-up, $t(10) = 2.91$, $p = .015$, $d_z = .88$.

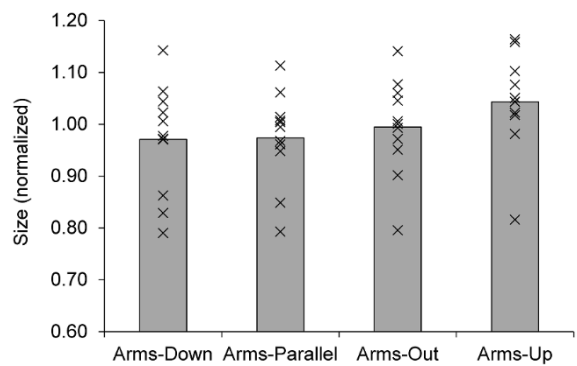

Figure 2. Sample mean (*bars*) and individual participant ( x ) normalized perceived size as a unitless ratio with respect to the physical dimensions (<1 indicates a smaller estimate than the veridical size) as a function of goalkeeper configuration (arms-down, arms-parallel, arms-out, arms-up).

For horizontal position, there was no significant effect of configuration, $F(3,30) = 1.97$, $p = .14$, $\eta_p^2 = .16$. Polynomial contrasts indicated only a significant cubic trend, $F(1,10) = 5.55$, $p = .040$, $\eta_p^2 = .36$, where the ball position was widest from the mid-line for arms-out (see Figure 3a). Meanwhile, the further t-test indicated no significant difference in the horizontal ball position between arms-down and arms-up, $t(10) = 1.09$, $p = .30$, $d_z = .33$.

For vertical position, there was a significant effect of configuration, $F(3,30) = 3.08$, $p = .042$, $\eta_p^2 = .24$. Polynomial contrasts indicated a significant linear trend, $F(1,10) = 10.31$, $p = .009$, $\eta_p^2 = .51$, where the ball position became progressively lower down from head-height when the arms were fanned higher up, including for arms-out and arms-up (see Figure 3b). Meanwhile, the further t-test indicated a significantly higher ball position for the arms-down compared to arms-up, $t(10) = 3.87$, $p = .003$, $d_z = 1.17$.

The correlation analyses revealed no significant relation between the perceived size and horizontal, $r = .35$, $p = .29$, nor vertical, $r = -.015$, $p = .97$, ball positions (see Figure 4).

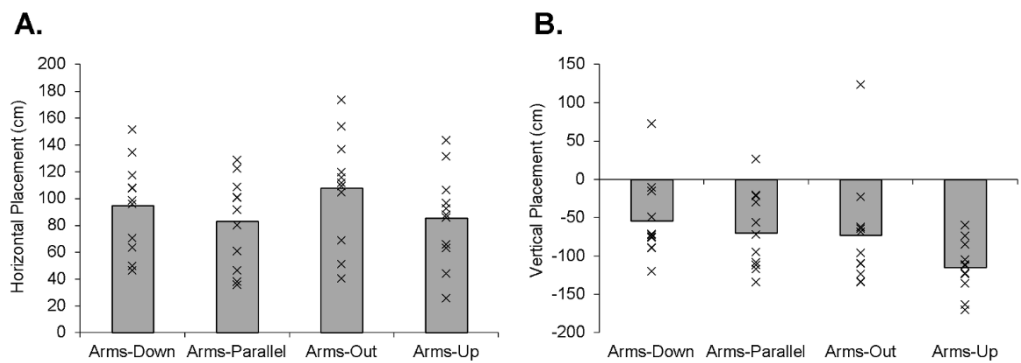

Figure 3. Sample mean (*bars*) and individual participant ( x ) horizontal (A) and vertical (B) ball position (cm) scores with respect to goalkeeper mid-line and head-height (<0 indicates below head-height), respectively, and as a function of goalkeeper configuration (arms-down, arms-parallel, arms-out, arms-up).

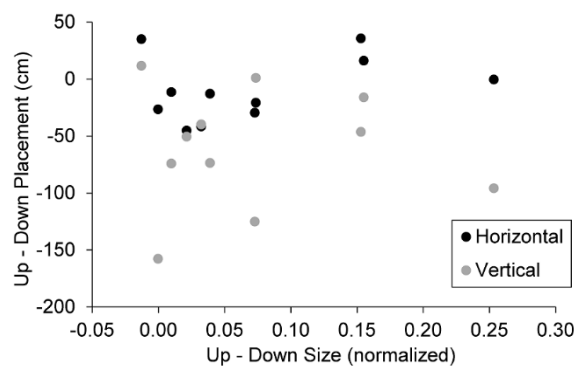

Figure 4. Relation between the mean normalized perceived size and ball position differences in arms-up and -down configuration (negative values indicate a lower magnitude for arms-up compared to arms-down). Horizontal and vertical positions indicated within the inset legend.

**Discussion**

The present study aimed to examine the influence of goalkeeper configurations, and more specifically, the potential illusory effects on penalty kicks. Participants completed perceptual and penalty kick tasks, where they had to indicate the perceived size of the goalkeeper height and execute penalty kicks toward a goalkeeper, respectively. The goalkeeper in-question adopted different configurations including arms-down, arms-parallel, arms-out and arms-up. Based on similar previous studies, we predicted an illusion-induced bias, where the kicks would end further away for the perceptually larger arms-up configuration, and perhaps closest for the perceptually smaller arms-down configuration. However, the findings indicated that despite a clear difference in the perceived size, which was consistent with the hypothesized direction of the illusion, there was no such effect on ball position. That is, the horizontal position was furthest away for the arms-out configuration with a limited difference between arms-down and arms-up, while the vertical position was progressively lowered when the arms transitioned from the down to up configurations. Likewise, there was no evidence of a correlation between the participants' perceived size of the goalkeeper and their subsequent kick position.

These findings may once more advocate for the functionally distinct visual neural pathways (Milner & Goodale, 1995; Glover, 2004; van der Kamp et al., 2008). Namely, the ventral pathway is suggested to compute the allocentric coordinates primarily for perception. In this instance, it may pertain to the shift in orientation of the arms with respect to the upper-body of the goalkeeper (0-135°), which gives rise to an illusion of a perceived size difference.

On the other hand, the dorsal pathway is suggested to compute the egocentric coordinates primarily for visually-guided action. This feature may pertain to the perceived position of the arms and veridical height of the goalkeeper independent of each other, which would manifest in a limited illusory effect.

Thus, at first glance, it would appear that the present study is in near opposition to previous other studies that have also examined the influence of goalkeeper configuration including the amputated version of the Müller-Lyer illusion (van der Kamp & Masters, 2008; Shim et al., 2014). For example, it was previously shown that the ball ended further away for the arms-up compared to arms-out configuration (van der Kamp & Masters, 2008), while another study indicated a pattern of ball positions that were almost entirely consistent with the direction of the illusion (Exp. 1-3; Shim et al., 2014). However, these sorts of discrepancies have been more aptly explained by subtle variations between studies including their choice of surrounding visual context and task constraints (for further details, see Bruno et al., 2008; Roberts et al., 2021). Along these lines, we may consider the potential influence of the adjacent goalposts independent of the illusory context related to the goalkeeper configuration. When there are no goalposts and no target goal to aim for, then the relative metrics from the goalkeeper configuration that manifest in an illusion may subsequently become accentuated (Shim et al., 2014). Alternatively, when goalposts are present and there is a clear indication of the target goal, then separate allocentric information pertaining to the relative location of the goalkeeper with respect to the goalposts may become available, and effectively cause any illusion to be overturned. Consistent with this logic are the findings from the present study, where the ball tended to end further away and lower down for arms-out and arms-up configurations, respectively; and thus clearly went for goal and avoided any tight areas of proximity between the goalkeeper and goalposts (for similar findings, see Masters et al., 2007; Weigelt & Memmert, 2012; Noël et al., 2016).

Moreover, we also highlight the potential role of the required action itself, including the use of ball-throwing from previous studies (van der Kamp & Masters, 2008; Shim et al., 2014), as opposed to the actual penalty kicks of the present study. With regard to the latter, penalty kicks typically feature an extended period of gaze fixation in order to preview the goal/goalkeeper followed by a gaze shift toward the ball for the penultimate approach and final kick (Wood & Wilson, 2010a). As a result, the allocentric coordinates that are closely attributed to any misperceived size of the goalkeeper may become extinguished by the time individuals arrive at the ball to kick it (Westwood & Goodale, 2003; see also, Elliott & Madalena, 1987; Elliott & Calvert, 1988).

Upon reflection, the present study comprises perception and action tasks that arguably contend with separate sets of visual cues at any one time (for a similar argument, see Smeets & Brenner, 2001; Smeets et al., 2002). To elucidate, the perceptual illusion task required participants to explicitly estimate the perceived size of the goalkeeper, and the penalty kick task alternatively required participants to kick the ball inside the goal while keeping it away from perceived reach of the goalkeeper. Hence, each of the tasks were not necessarily comparable, and there could not reasonably be any close relation between the measures of perceived size of the goalkeeper and ball position within the penalty kicks (see Figure 4). With this in mind, perhaps it remains of interest whether the illusory context of goalkeeper configurations could still somehow influence penalty kicks. For example, future research may manipulate the instructions and subsequent areas of gaze fixation to help emphasise the allocentric cues that are specific to the goalkeeper ('goalkeeper-dependent'), as opposed to the target goal ('goalkeeper-independent') (Wood & Wilson, 2010b; see also, Navarro et al., 2013), which could conceivably promote an illusion-induced bias. In addition, we might also consider the time spent in preparation and/or execution of penalty kicks (e.g., Mendoza et al., 2006; Roberts et al., 2017), whereby prolonging the preparation (e.g., >2 s) could cause the interim storage of the goalkeeper configuration to decay, and thus limit an illusion-induced bias.

In summary, the present findings indicate that while goalkeeper configurations can effectively adopt a perceptual illusion to elicit a misperception in size, it does not necessarily effect penalty kicks in the same way. That is, the kicks tended to be directed in such a way that they could positively avoid the goalkeeper, while still reaching the target goal. We suggest that the current penalty kicks were mostly influenced by an intuitive attempt to avoid tighter areas of proximity between the goalkeeper and goalposts, where there could possibly be a reduced chance of scoring (i.e., save or miss). However, the present findings and related interpretation should be treated with a degree of caution owing to certain limitations. Namely, there were potentially confounding background visual characteristics, as well as a separation in the visual information (e.g., goalkeeper vs. goal) and measurement sensitivity (e.g., refined pixelated size differences vs. coarse end ball position differences) between each of the perceived size and penalty kick tasks. In addition, despite the apriori power analysis, the comparatively small sample size ($n = 11$) may question the generalisability of this study. These issues, along with the fore mentioned study possibilities, may lay the groundwork for continued research on the influence of goalkeeper configurations within soccer penalty kicks.